# Complex Signal Processing for Coriolis Mass Flow Metering in Two-Phase Flow

Ming Li, Manus Henry

*Abstract*— This paper presents a new signal processing method based on Complex Bandpass Filtering (CBF) applied to the Coriolis Mass Flowmeter (CMF). CBF can be utilized to suppress the negative frequency component of each sensor signal to produce the corresponding analytic form with reduced tracking delay. Further processing of the analytic form yields the amplitude, frequency, phase and phase difference of the sensor signals. In comparison with previously published methods, CBF offers short delay, high noise suppression, high accuracy and low computational cost. A reduced delay is useful in CMF signal processing especially for maintaining flowtube oscillation in two/multi-phase flow conditions. The central frequency and the frequency range of the CBF method are selectable so that they can be customized for different flowtube designs.

## I. INTRODUCTION

The Coriolis Mass Flowmeter (CMF) provides direct and accurate measurement of the mass flow and (usually) the density of a single-phase fluid. This capability has led to its widespread industrial application over several decades [1]. A CMF has two primary components: the flowtube, with typically two velocity sensors, and at least one driver to induce flowtube oscillation; and the transmitter, an embedded system for signal processing, measurement and control. Recently, CMF applications have been extended to two-phase, gas/liquid mixtures [2]–[4]. The signal processing requirements for CMF are challenging, especially in two-phase conditions [5], [6], entailing the tracking of amplitude, frequency, phase and phase difference simultaneously for the essentially sinusoidal signals.

For example, Fig. 1 shows a typical pair of sensor signals when monitoring the flow of a single phase liquid, while Fig. 2 shows the corresponding signals when a high level of gas is included in the liquid flow to create a two-phase mixture. The flowtube damping is constantly changing, as is the mixture flow rate and density. These changes result in much greater variation in frequency, amplitude and phase difference in the corresponding CMF sensor signals. Accordingly, signal processing for CMF in two-phase conditions are more challenging than for a single phase fluid.

One aspect of CMF operation less frequently discussed in the literature is the time delay between sensor input and driver output. Since a primary requirement for CMF operation is to maintain flowtube oscillation, usually at a fixed amplitude [7], a synthesized driver signal is generated to be in phase with a sensor signal (or often the sum – and hence mid-phase – of the two sensor signals). In this process, time delay plays an important role in determining the quality of the amplitude control. With single phase flow, the time delay in the flow tube control system is less important as the sensor signals are inherently more stable. But with two-phase flow, as illustrated in Fig. 2, time delay in the flowtube control system becomes a more significant factor in regulating the rapidly changing oscillation. A further problem caused by two-phase flow is the rise in mechanical damping on the flowtube. As illustrated by comparing Figs. 1 and 2, higher damping often results in a lower amplitude of oscillation, as the maximum drive energy is usually limited. With lower and varying amplitude, and higher levels of background noise, the Signal-to-Noise Ratio (SNR) is significantly reduced in two-phase flow, and so noise suppression is also important.

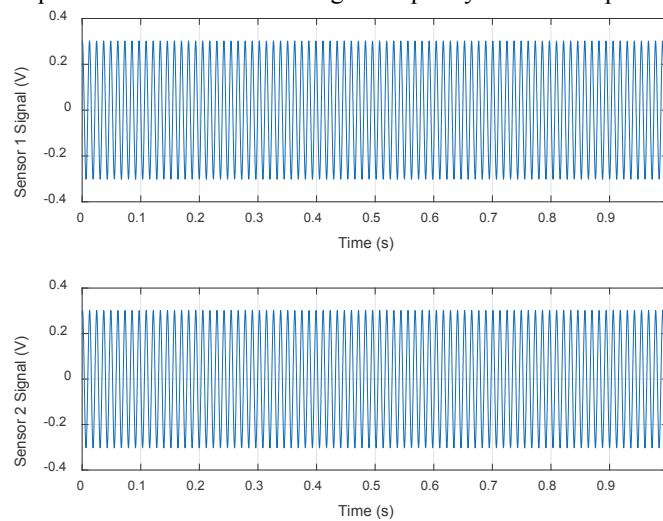

Fig. 1.  CMF sensor signals with single phase flow

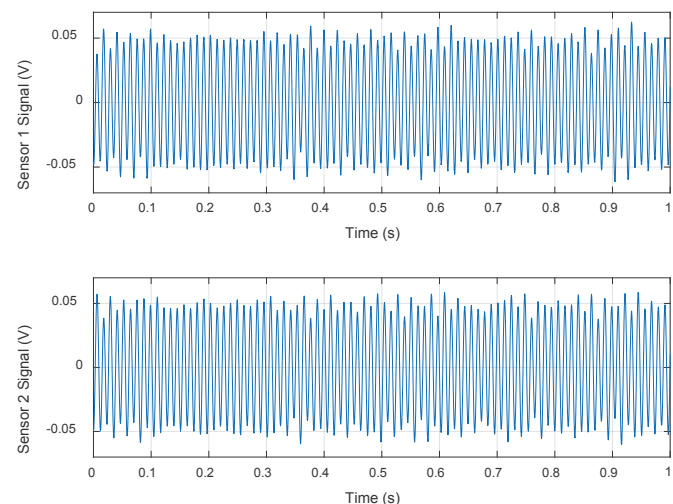

Fig. 2.  CMF sensor signals with two-phase flow

With a digital drive, time lags occur between the sensor and driver signals due to Analog-to-Digital Converter (ADC) and Digital-to-Analog Converter (DAC) operations, signal processing and phase synchronization. Fig. 7 in [8] shows a typical timing diagram for the drive generation process. Note that a large proportion of this delay is due to the measurement process itself, i.e. the extraction of frequency, phase and amplitude information from the sensor signal. Minimizing this delay improves the responsiveness of the amplitude control, and indeed the flow measurement itself, an increasingly important issue in a number of applications [9], [10].

Current methods for CMF signal processing include the Hilbert Transform (HT) [11]–[14], dual quadrature demodulation [15], digital correlation [16], and a combination of the Adaptive Notch Filter (ANF) for frequency tracking together with the Discrete-Time Fourier Transform (DTFT) for amplitude and phase calculation [17]–[19]. These methods work well under single-phase, steady flow conditions since all the parameters are relatively stable. But there has been little previous discussion in the literature about dynamic response and performance in two-phase flow conditions [5]. Prism signal processing [20] is a new technique based on recursive FIR filtering: it has been applied in Coriolis metering to monitor automobile fuel injection pulses as short as 1 ms [9], thus demonstrating the potential for fast dynamic response. However, Prism signal processing has not yet been applied to the two-phase flow problem.

In summary, accurately tracking CMF sensor signals, when subject to time-varying amplitude and frequency, and while minimizing measurement delay, are important challenges for the next generation of two-phase flow capable CMF.

Based on these requirements, three complex signal processing methods – Complex Bandpass Filter (CBF), Complex Notch Filter (CNF) and Complex Bandpass Filter with Complex Notch Filter (CBF-CNF), have been developed. The key idea is to suppress the negative frequency of the sensor signal in order to generate the corresponding analytic form, from which each required parameter value can be calculated. Also, due to their natural bandpass property, the new techniques need not require pre-filtering, thus reducing time delay and introducing an ability to suppress noise.

In section II, the complex signal processing is described for each of the CBF, CNF and CBF-CNF methods. In section III, the new techniques are compared with existing methods. Simulation results with quantified error performances are presented. In Section IV, the complexity of each algorithm is compared. Finally, in section V, the findings are summarized.

## II. COMPLEX SIGNAL PROCESSING

### A. Complex bandpass filter

The CBF is derived from a conventional low-pass filter design (for example an IIR elliptic filter) through the application of a complex shift factor $e^{j\theta}$ to the filter coefficients. Define the original IIR filter transform function as:

$$H_r(z) = \frac{\sum_{m=0}^{P} b_m z^{-m}}{\sum_{n=0}^{Q} a_n z^{-n}} \quad (1)$$

where $P$ and $Q$ are the numerator and denominator order, and $b_m$ and $a_n$ are the filter coefficients. If $Q = 0$, the filter is Finite Impulse Response (FIR); otherwise, it is Infinite Impulse Response (IIR).

Applying the complex shift factor $e^{j\theta}$ to $H_r(z)$:

$$H_r(z) \xrightarrow{z^{-1}=e^{j\theta}z^{-1}} H_c(z) = \frac{\sum_{m=0}^{P} b_m e^{j\theta m} z^{-m}}{\sum_{n=0}^{Q} a_n e^{j\theta n} z^{-n}} \quad (2)$$

The numerator and denominator orders $P$ and $Q$ are unchanged, while the filter coefficients become $b_m e^{j\theta m}$ and $a_n e^{j\theta n}$.

The complex shift rotates the zeros and poles of the original filter by an angle $\theta$ in the z-plane in the anti-clockwise direction. Fig. 4 shows the resulting zero and pole rotation for an exemplary 5th order elliptic filter.

The rotation only changes the angle of the poles and zeros in the z-domain, while the radius and relative positions remain the same. Accordingly, the effect on the filter properties is to induce a shift of $\theta$ radians/sample in the magnitude response, as shown in Fig. 5, converting the original low pass filter into a bandpass filter.

When a real signal is passed through the CBF, the negative frequency component is filtered out, resulting in a single-sided analytic form. The filtering process in the frequency domain is shown in Fig. 6, where the original input is a real double-sided sinusoidal signal with frequency $\omega_0$.

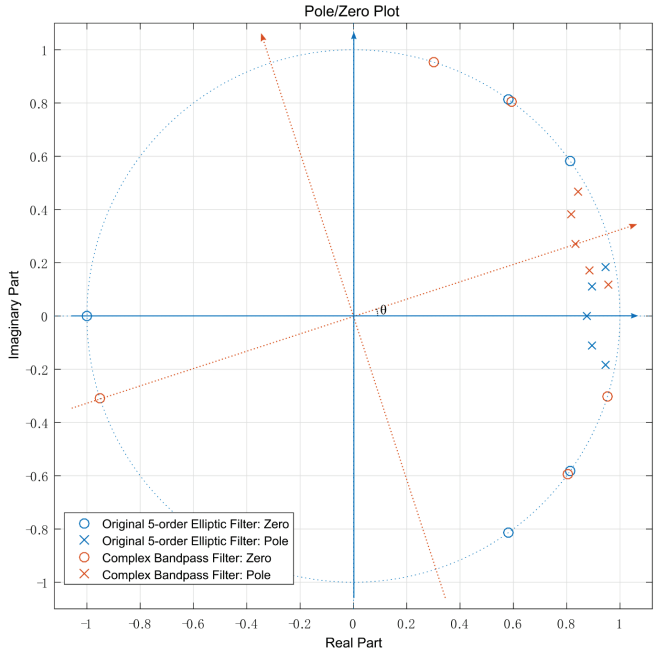

Fig. 4. Zeros and poles rotation for CBF

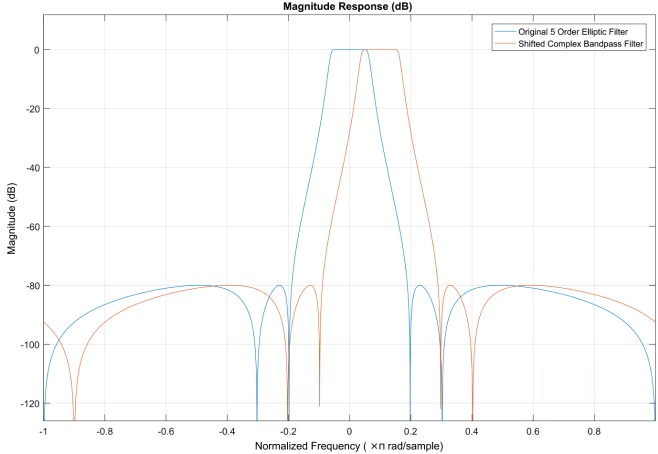

Fig. 5. Original elliptic filter and shifted CBF magnitude response

After getting the analytic form of the filtered signal, the calculation of amplitude, frequency, phase are similar to those used in the Hilbert transform method [11]. For CFM signal processing, where two sensor signals must be tracked, the phase difference is readily calculated from the phases of the individual signals. Assume the CFM sensor signals are:

$$x_1 = A_1 \sin(\omega t + \phi/2)$$
$$x_2 = A_2 \sin(\omega t - \phi/2) \quad (3)$$

where $A_1$ and $A_2$ are the amplitudes of the two sensor signals, $\omega$ is their common frequency, and $\phi$ is the phase difference between them. Passing the signals through identical CBF filters, complex analytic signals are obtained as follows:

$$x_{1a} = A_1[\cos(\omega t + \phi/2) + i\sin(\omega t + \phi/2)] = A_1 e^{j(\omega t + \phi/2)}$$
$$x_{2a} = A_2[\cos(\omega t - \phi/2) + i\sin(\omega t - \phi/2)] = A_2 e^{j(\omega t - \phi/2)} \quad (4)$$

The amplitudes can be obtained by taking conjugate terms:

$$x_{1a} \times \overline{x_{2a}} = A_1 e^{j(\omega t + \phi/2)} \times A_2 e^{-j(\omega t - \phi/2)} = A_1 \times A_2 e^{j\phi} \quad (5)$$

For brevity we assume here that the two input signals have the same amplitude ($A_1 = A_2 = A$); this is a reasonable first approximation for the CMF, and further computational steps

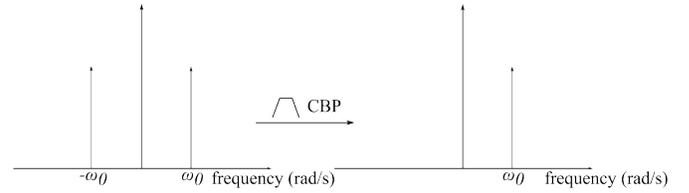

Fig. 6. Filtering process of CBF

can be taken to account for any sensor imbalance. Equation (5) is thus simplified to:

$$x_{1a} \times \overline{x_{2a}} = A^2 e^{j\phi} \quad (6)$$

The phase difference may now be calculated using:

$$\phi = \arg(x_{1a} \times \overline{x_{2a}}) \quad (7)$$

while the amplitude can be obtained using:

$$A_1 = A_2 = |x_{1a}| = |x_{2a}| \quad (8)$$

The frequency may be derived from the change of phase over a number of samples. For example, consider adjacent samples from the first sensor signal:

$$\overline{x_{1a}(n-1)} \times x_{1a}(n) = Ae^{-j(\omega t_{n-1} + \phi/2)} \times Ae^{j(\omega t_n + \phi/2)}$$
$$= A^2 e^{j(wt_n - wt_{n-1})} \quad (9)$$

followed by:

$$\omega t_n - \omega t_{n-1} = \arg(\overline{x_{1a}(n-1)} \times x_{1a}(n)) \quad (10)$$

Converting into Hz:

$$f = \frac{(\omega t_n - \omega t_{n-1}) \times F_s}{2\pi} \quad (11)$$

where $F_s$ is the sampling frequency. Using adjacent samples to calculate frequency provides a short response time but is susceptible to noise; comparing the phase shift over a larger number of samples is equally possible, leading to a tradeoff between measurement noise and response time.

By selecting the 'shift' frequency and passband of the CBF, the range of frequencies that may be tracked can be matched to the resonant frequency range of the corresponding flowtube.

### B. Complex notch filter

Another way to generate the analytic form of a real sinusoidal signal is to apply a Complex Notch Filter (CNF) to filter out the negative frequency component. CNF is advantageous in that the tracking delay may be close to zero at the retained positive frequency. The CNF is designed such that the notch frequency is located at the negative frequency of the real sensor signal. For a CFM, where a range of frequencies may be output from a flowtube, a bandstop filter can be created by shifting a high-pass filter to the negative side of the frequency domain. In this case, the transfer function of the CNF follows equation (2) but the shift angle becomes $-\theta$ which means zeros and poles rotates clock-wise z-plane. The transfer function becomes:

$$H_r(z) \xrightarrow{z^{-1} = e^{-j\theta} z^{-1}} H_c(z) = \frac{\sum_{m=0}^{P} b_m e^{-j\theta m} z^{-m}}{\sum_{n=0}^{Q} a_n e^{-j\theta n} z^{-n}} \quad (12)$$

Fig. 7 shows zeros and poles from a 5th-order elliptic high-pass filter rotated to form a CNF with notch at $\omega_c$ = -0.1 rad/s.

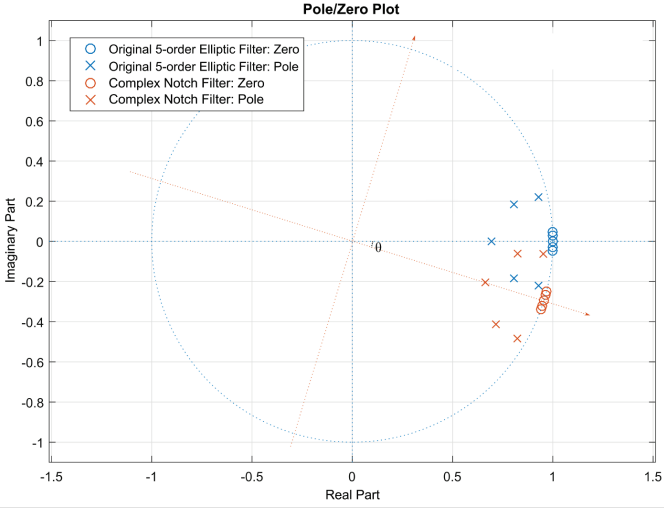

Fig. 7. Zeros and poles rotation for CNF

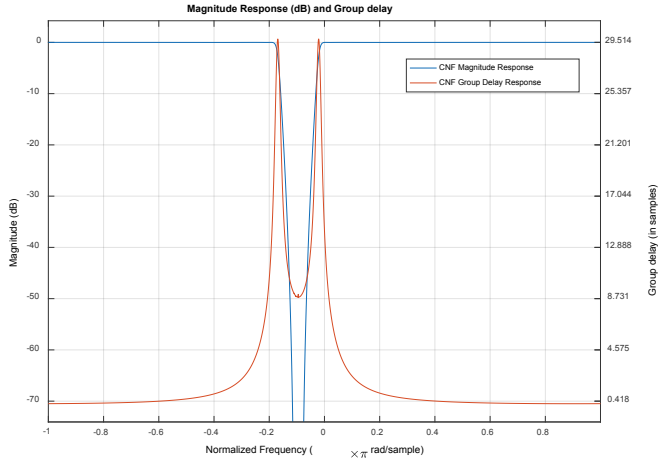

Fig. 8. Group delay of 5th order CNF

From Fig. 8, the CNF's group delay at the corresponding positive frequencies (around 0.1 rad/s) is close to 2 samples. Generally, it would be expected that the CNF tracking delay should be significantly reduced compared to that of the CBF. A comparison via simulation is given in Section III D below.

CNF may be extended to form a comb filter to remove harmonic noise, e.g. a mains frequency (50 or 60 Hz) along with its harmonics. By adding notches at a series of harmonics, the CNF can be turned into a comb structure while still retaining a small delay at the positive tracking frequency.

One disadvantage of CNF is that it provides limited noise reduction compared with the bandpass characteristic of the CBF. This occurs because only the notched negative frequency is attenuated, while all other noise components are passed through into the sinusoid parameter tracking calculations.

### C. Complex bandpass filter with complex notch filter

While CBF has good noise suppression but a relatively large tracking delay, CNF is sensitive to noise but has a short tracking delay. This suggests it may be possible to achieve a trade-off between noise performance and tracking delay for complex signal processing.

Such a trade-off may be achieved by cascading CBF and CNF together (CBF-CNF). In this combined approach, the design of the CBF section need not be as constrained in terms of passband, roll-off, and stop band attenuation, potentially reducing tracking time delay and improving accuracy during dynamic change. Here we combined a 3rd-order CBF with $4^{th}$-order CNF to form CBF-CNF for use in simulations.

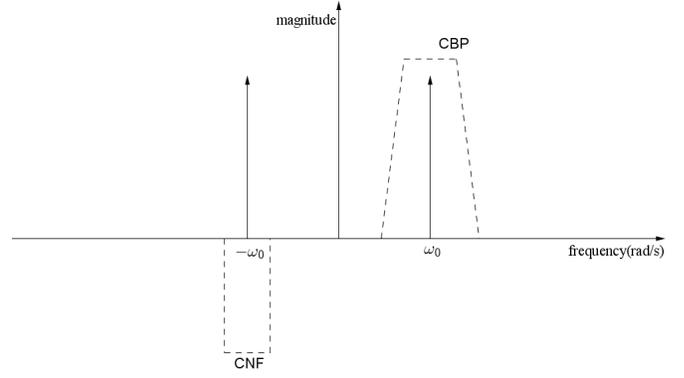

Fig. 9. Filtering process of CBF-CNF

### III. SIMULATION

In order to evaluate their tracking performances, the CBF and CBF-CNF techniques are compared with established sinusoidal tracking methods in a simulation of two-phase flow conditions, including empty-full-empty batching.

#### A. Existing methods

Two current techniques are used in simulation: ANF for frequency tracking combined with the DTFT for amplitude and phase calculation [17]–[19], which is denoted as DTFT (ANF) here; and the Hilbert Transform (HT) [11]–[14]. The sampling frequency $f_s$ is 2 kHz, which is sufficiently higher than the CMF vibrating frequency at around 100 Hz.

For DTFT (ANF), we follow the technique described in [12] which uses the Steiglitz-McBride ANF (SMM-ANF) structure. The transfer function is

$$H(z) = \frac{\hat{A}_n(z^{-1})}{\hat{A}_n(\rho z^{-1})} = \prod_{k=1}^{m} \frac{1+\hat{\alpha}_k(n)z^{-1}+z^{-2}}{1+\rho\hat{\alpha}_k(n)z^{-1}+\rho^2 z^{-2}} \quad (13)$$

where $m$ is the trap number i.e. the number of peak frequencies to be tracked. Since the CMF signal has a single dominant frequency, $m=1$. $\rho$ is the pole contraction factor which determines the bandwidth of the ANF, given by:

$$BW = 2\cos^{-1}\left(\frac{2\rho}{1+\rho^2}\right) \quad (14)$$

The weight coefficient $\hat{\alpha}_k(n) = -2\cos\hat{\omega}_k(n)$, with $\hat{\omega}_k(n)$ the notch frequency estimate of the input frequency, is adjusted by a Recursive Mean Square (RMS) algorithm:

$$\begin{aligned}
\hat{\alpha}_k(n+1) &= \hat{\alpha}_k(n) - P(n)\psi(n)e_s(n) \\
\psi(n) &= \frac{\partial e_s(n)}{\partial \hat{\alpha}_k(n)} = \frac{y(n-1)-\rho e_s(n-1)}{1+\rho\hat{\alpha}_k(n)z^{-1}+\rho^2 z^{-2}} \\
P(n) &= \frac{(1-\lambda)P(n-1)}{\lambda+\psi^2(n)P(n-1)}
\end{aligned} \quad (15)$$

where $\psi(n)$ is the gradient function, $P(n)$ represents the covariance parameters, $\lambda$ is the forgetting factor, and $e_s(n)$ represents the output of ANF.

When implementing the SMM-ANF algorithm in the simulation described below, the initial values are:

$$\rho = 0.9, P(1) = P(2) = 100, \lambda = 0.9 \quad (16)$$

Then after obtaining frequency, the amplitude and phase difference can be calculated using a recursive DTFT:

$$DFT_R(n, \omega_k) = [x(n+N/2-1) - x(n-N/2-1)](-1)^k e^{\frac{j2\pi\omega_k}{N}}$$
$$+ DFT_R(n-1, k) e^{\frac{j2\pi\omega_k}{N}} \quad (17)$$

where $x(n)$ is the input signal function, $N$ is the window length, $k$ is the sampling point ranging from $0, 1, \cdots, N/2-1$ and $\omega_k$ is the estimated frequency from ANF at $k^{th}$ sampling point.

For HT, we follow the method presented in [11] using Parks-McClellan FIR filter design method to form a $49^{th}$-order FIR filter. Passing the input signal through HT generates an imaginary signal orthogonal to the original real input signal, and so the analytic form is obtained.

### B. Two-phase flow simulation

In order to simulate the sensor signals arising in two-phase flow conditions, a Random Walk Model (RWM) was proposed by Tu *et al.* [12]. However, the technique places no limits on the instantaneous change of the parameter values. In a real CMF, the rate of change of each parameter is physically limited due to mechanical inertia, limited fluid velocities and so on. Here we introduce filtering of the parameter changes in order to create a more realistic simulation – Modified Random Walk Model (MRWM). We define parameters as follows:

$$y_1(n) = A(n)\sin[\omega(n)n + \phi(n)/2] + \sigma_{e1} \cdot e_1(n)$$
$$y_2(n) = A(n)\sin[\omega(n)n - \phi(n)/2] + \sigma_{e2} \cdot e_2(n)$$
$$A(n) = \frac{(A_f(n) - \min(A_f(n))) \times (A_{\max} - A_{\min})}{\max(A_f(n)) - \min(A_f(n))} + A_{\min}$$
$$\omega(n) = \frac{(\omega_f(n) - \min(\omega_f(n))) \times (\omega_{\max} - \omega_{\min})}{\max(\omega_f(n)) - \min(\omega_f(n))} + \omega_{\min} \quad (18)$$
$$\phi(n) = \frac{(\phi_f(n) - \min(\phi_f(n))) \times (\phi_{\max} - \phi_{\min})}{\max(\phi_f(n)) - \min(\phi_f(n))} + \phi_{\min}$$
$$A_f(n) = h_A(n) * e_A(n)$$
$$\omega_f(n) = h_\omega(n) * e_\omega(n)$$
$$\phi_f(n) = h_\phi(n) * e_\phi(n)$$

where $e_1(n)$ and $e_2(n)$ are un-correlated white noise sequences, $e_A(n)$, $e_\omega(n)$ and $e_\phi(n)$ are uniformly distributed random noise signals in the interval (-1,1), and $\sigma_{e1}$, $\sigma_{e2}$ are gain factors for the input noise. $A_{\max}$, $A_{\min}$ are the upper and lower limits for the time-varying amplitude, and where $\omega_{\max}$, $\omega_{\min}$, $\phi_{\max}$ and $\phi_{\min}$ are the corresponding limits for frequency and phase difference respectively. $h_A(n)$, $h_\omega(n)$ and $h_\phi(n)$ are low-pass filters to limit the rate of change of parameter values and the '*' operator is the time domain convolution process for filtering.

The bound-limited amplitude, frequency and phase difference values are generated via uniform random noise processes and passed through a low-pass filter to constrain the rate of change. Table I shows parameter values for MRWM used for simulation in this paper. Then, based on the filtered parameter values, the two sensor signals are generated. Fig. 10 compares the output of MRWM with that of RWM.

From Fig. 10, both the RWM and MRWM range over the desired parameter values in a random fashion. However, MRWM excludes high frequency variations which are physically unrealistic, but which will influence the error statistics and hence the performance evaluation of the measurement techniques.

TABLE I
PARAMETER VALUES FOR MRWM

| Parameters | Values |
|---|---|
| Sampling frequency ($f_s$) | 2 kHz |
| Low-pass filter cut-off frequencies ($f_A^s$, $f_\omega^s$, $\omega_\phi^s$) | 6 Hz |
| Range for amplitude | $A_{\min}$ = 0.05 V, $A_{\max}$ = 0.3 V |
| Range for frequency | $f_{\min}$ = 85 Hz, $f_{\max}$ = 100 Hz |
| Range for phase difference | $\phi_{\min}$ = 0°, $\phi_{\max}$ = 4° |

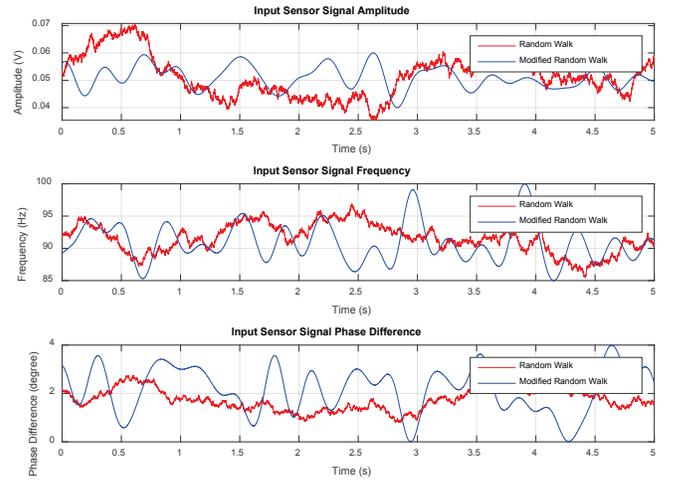

Fig. 10. Comparison between RWM and MRWM

### C. Simulation Results

Based on signals generated using MRWM, simulations of CBF and CBF-CNF have been carried out alongside the DTFT (ANF) and HT methods. Fig. 11 shows typical sensor data supplied to each of the tracking algorithms over 1 s period, which is similar to the real sensor data shown in Fig. 2.

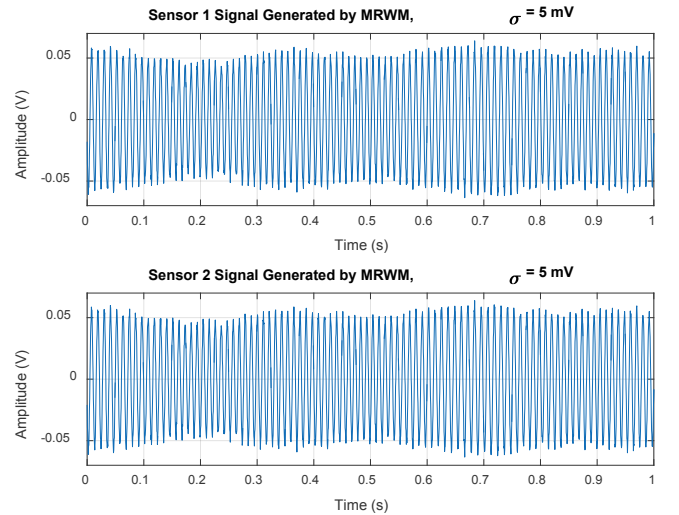

Fig. 11. Sensor signals generated by MRWM

TABLE II
TWO-PHASE SIMULATION QUANTIFIED TRACKING PERFORMANCE

| RMSE in Noise Free Experiment | | | | |
|---|---|---|---|---|
| Method and Parameter | DTFT (AANF) | Hilbert | CBF | CBF-CNF |
| Amplitude (V) | 9.580e-4 | 6.786e-4 | 4.574e-4 | 2.983e-4 |
| Frequency (Hz) | 1.365e+0 | 5.294e-1 | 3.562e-1 | 2.052e-1 |
| Phase Diff (°) | 2.458e-1 | 1.404e-1 | 9.551e-2 | 6.076e-2 |
| RMSE in $\sigma$ = 5 mV Experiment | | | | |
| Method and Parameter | DTFT (AANF) | Hilbert | CBF | CBF-CNF |
| Amplitude (V) | 2.512e-2 | 2.903e-3 | 1.123e-3 | 1.031e-3 |
| Frequency (Hz) | 1.567e+0 | 2.280e+1 | 8.991e-1 | 1.185e+0 |
| Phase Diff (°) | 3.286e-1 | 1.944e-1 | 1.077e-1 | 7.756e-2 |

As discussed above, CNF used in isolation is sensitive to noise, and so it is not included in this simulation study. Accordingly, the CBF, CBF-CNF, DTFT (ANF) and HT methods are tested using the time-varying input shown in Fig. 11. In further simulations, white noise with standard deviation $\sigma$ is added to the sensor signals. With $\sigma$ = 5 mV and a time varying amplitude the average SNR is approximately 20 dB.

Figs. 13-15 show the methods' tracking performance in noise-free conditions while Figs. 16-18 show the corresponding performance with the addition of noise. From the figures, CBF and CBF-CNF outperform DTFT (ANF) and the Hilbert method, having smaller tracking delay and a better dynamic response. CBF-CNF also has the smallest tracking delay and shows better tracking in the noise-free case. For the noisy case, since CBF is designed to have deeper stopband attenuation, the tracking result is smoother especially for frequency.

### D. Quantified error performance

To evaluate each method's performance numerically, Table 2 shows the Root Mean Squared Error (RMSE), calculated by:

$$RMSE = \sqrt{\frac{1}{n}\sum_{i=1}^{n}(\hat{Y}(i) - Y(i))^2} \quad (19)$$

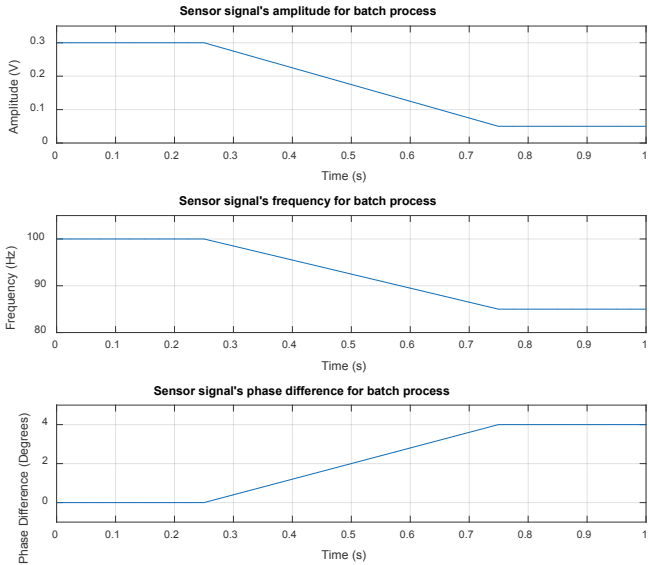

Fig. 12. Batch process simulation sensor signal parameters

TABLE III
BATCH PROCESS SIMULATION QUANTIFIED TRACKING PERFORMANCE

| Method and Parameter | DTFT (AANF) | Hilbert | CBF | CNF | CBF-CNF |
|---|---|---|---|---|---|
| Amplitude (V) | 2.64e-5 | 2.12e-5 | 9.28e-6 | 1.09e-6 | 2.02e-6 |
| Frequency (Hz) | 7.41e-1 | 6.68e-2 | 2.82e-3 | 3.35e-3 | 3.47e-3 |
| Phase Diff (°) | 9.87e-2 | 5.36e-3 | 2.48e-3 | 3.14e-4 | 5.30e-4 |
| Tracking Delay (ms) | 15 | 12.5 | 8.75 | 3.125 | 3.75 |

where $Y(i)$ and $\hat{Y}(i)$ are the true and estimated values. The results are shown in Table II for the noise free and $\sigma$ = 5 mV experiments.

From Table II, CBF-CNF performs best in the noise-free simulation and in the $\sigma$ = 5 mV simulation except for frequency tracking. This is because the frequency calculation is susceptible to high noise and requires deeper stop-band attenuation to track frequency well. However the CBF technique outperforms the existing methods especially in noisy conditions.

### E. Dynamic Performance

When comparing dynamic performance, one particularly challenging condition for maintaining good measurement tracking and flowtube control occurs during a rapid transition from empty to full or from full to empty. This condition typically arises in a batching operation, where the Coriolis meter starts and ends empty, and is required to report the total flow of liquid in the batch [9]. Such a transition, with the associated changes in mass flow, density and flowtube damping, is likely to result in simultaneous changes in frequency, amplitude and phase difference on the two sensor signals, over a period as short as 0.5 s.

All three new techniques have been tested alongside the DTFT (ANF) and HT methods in a start of batch simulation. Here an empty-to-full flowtube filling process is simulated via otherwise noise-free sensor signals where the common frequency drops linearly from 100 Hz down to 85 Hz, the amplitudes drops linearly from 0.3 V down to 0.05 V and the phase difference increases linearly from 0° up to 4°, all simultaneously over 0.5 s. The simulation sample rate is 2 kHz. Fig. 12 shows the batch process simulation input signals' frequency, amplitude and phase difference.

Table III shows the simulation results for each method with approximate tracking time delay. From the result, CBF, CNF and CBF-CNF all perform better than existing methods with less tracking delay.

## IV. COMPLEXITY

Further key aspects of algorithm assessment include the computational complexity and data storage requirements, as CMF measurement algorithms need to be implemented in embedded systems and to run in real-time. Table V compares the conventional and new techniques in terms of the static memory size required for buffering data and storing variables, and the number of additions (or subtractions) and multiplications (or divisions) required to process each new sample. The assessment of complexity includes the full calculation from pre-filtering through to the estimation of frequency, amplitude and phase difference. It excludes

TABLE V
COMPLEXITY OF EACH ALGORITHM

| Method | DTFT (AANF) | Hilbert | CBF | CNF | CBF-CNF |
|---|---|---|---|---|---|
| Additions | 21 | 99 | 21 | 21 | 25 |
| Multiplications | 53 | 108 | 26 | 26 | 30 |
| Static Storage (bytes) | 432 | 800 | 320 | 320 | 448 |

however the arctangent calculation used once per sample that is common to all the methods listed.

From Table V, it can be seen that the CBF, CNF and CBF-CNF techniques have relatively low computational requirements. This will support the use of the newly developed techniques in real-time, low cost, implementations.

## V. CONCLUSION

This paper has described three complex bandpass filtering methods and has applied them to CMF signal processing. The CBF can be derived from a simple low-pass filter with a selectable central frequency and bandwidth. The calculation is simple and computational cost is small. Due to the nature of bandpass filtering, the CBF can not only track amplitude, frequency and phase difference at the same time, but it also applies strong noise suppression, which is increasingly important in CMF applications. Additionally, CBF combined with CNF has been used to reduce tracking delay and improve accuracy. Simulation studies suggest the tracking performance of these two methods is generally superior to that of the DTFT (ANF) and Hilbert transform techniques. The computational cost and size is also less than that of the existing techniques.

A future publication will describe experimental results in which the real-time measurement and control performance of the CBF and CBF-CNF algorithms have been compared with that of a commercially applied algorithm over a range of two-phase flow conditions. These results will further demonstrate that CBF and CBF-CNF offer a good signal processing solution to meet future CMF signal processing requirements.

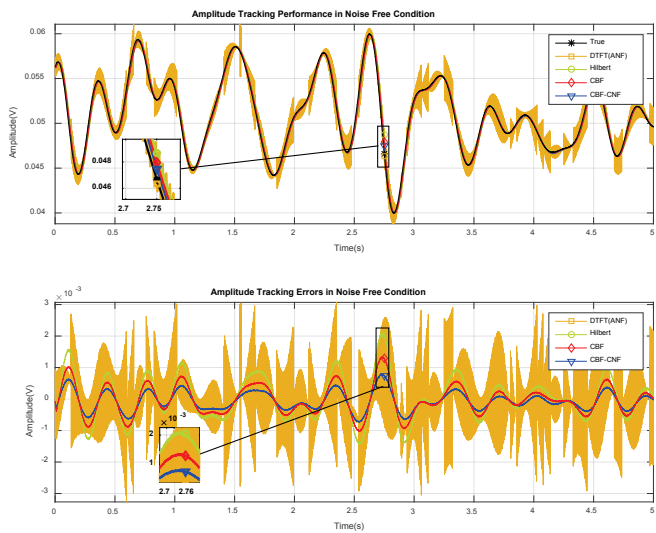
Fig. 13. Amplitude tracking performance in noise free conditions

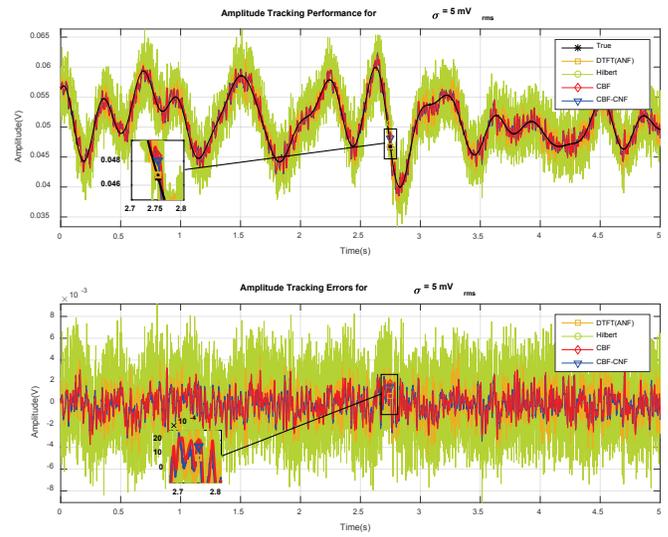
Fig. 16. Amplitude tracking performance for σ = 5 mV

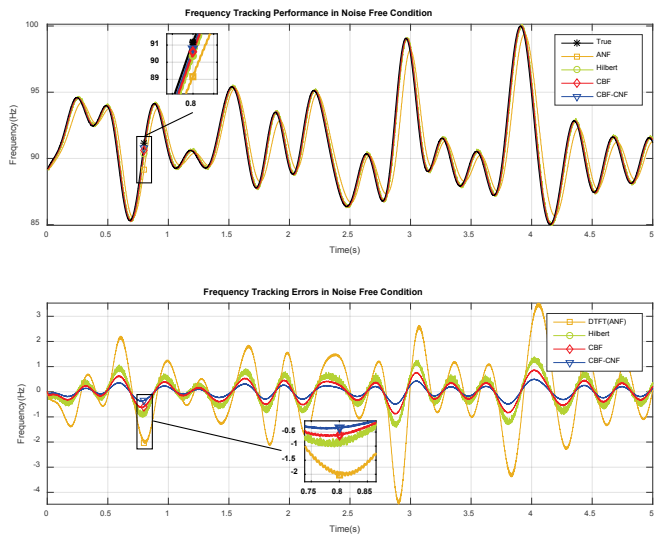
Fig. 14. Frequency tracking performance in noise free conditions

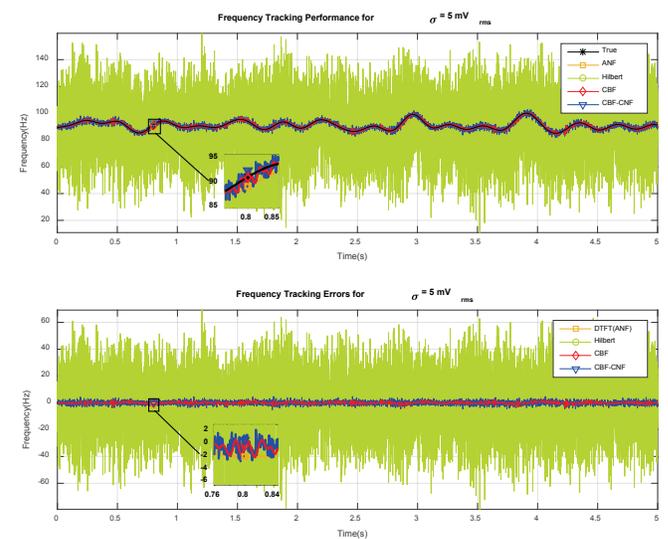
Fig. 17. Frequency tracking performance for σ = 5 mV

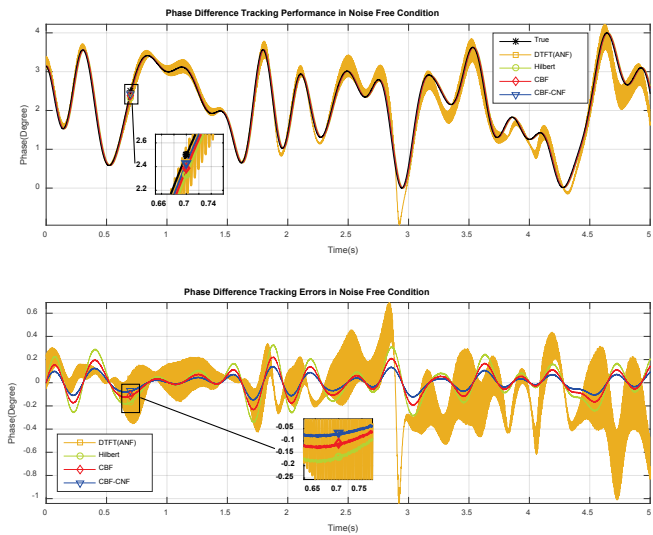
Fig. 15. Phase difference tracking performance in noise free conditions

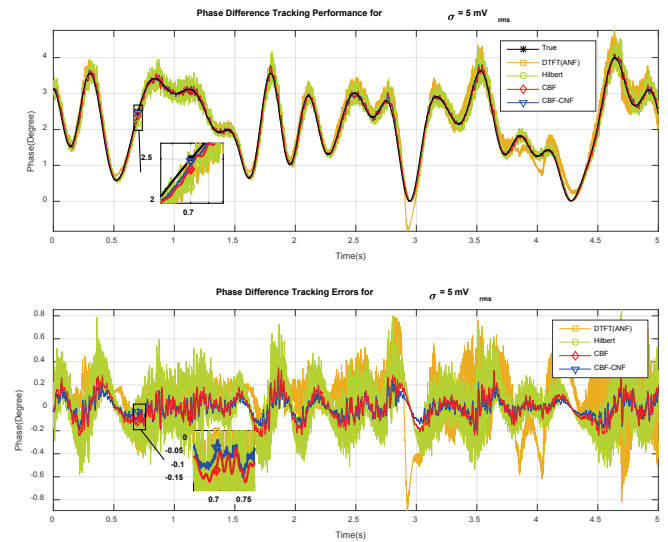
Fig. 18. Phase difference tracking performance for σ = 5 mV